\documentstyle[11pt,epsfig]{article}  
\begin{document} 

%%%%%%%%%%%%%%%%%%%%%%%% TITLE PAGE

\begin{titlepage} 
\begin{center}
{\Large \bf Jet quenching and high-$p_T$ azimuthal asymmetry~\footnote{ 
Contribution to the CERN Yellow Report on Hard Probes in Heavy Ion 
Collisions at the LHC.}  } 

\vspace{5mm}

{\it I.P.~Lokhtin$^a$, A.M.~Snigirev$^a$ and I.~Vitev$^b$} 

\vspace{2.5mm} 
    
$^a$ Institute of Nuclear Physics, Moscow State University, 119992,
Moscow, Russia \\ 
$^b$ Iowa State University, 12 Physics Bldg. A330, Ames, IA 50011, USA  

\vspace{5mm} 

\end{center}  

\begin{abstract} 
The azimuthal anisotropy of high-$p_T$ particle production in
non-central heavy ion collisions is among the most promising observables of
partonic energy loss in an azimuthally non-symmetric volume  of quark-gluon plasma. 
We discuss the implications of nuclear geometry for the models of partonic energy 
loss in the context of recent RHIC data and consequences  for observation 
of jet quenching at the LHC.  
\end{abstract}

\end{titlepage}   

\section{Introduction} 

In order to interpret data on nuclear collisions from current  experiments at 
the Relativistic Heavy Ion Collider (RHIC) and  future  experiments at the 
Large Hadron Collider (LHC), it is  necessary to have knowledge of the 
{\em initial conditions}. There are large  uncertainties in  the estimates of 
the initial produced  gluon density,  $\rho_g (\tau_0) 
\sim 15 - 50 / {\rm fm}^3$  in central  $Au+Au$ at  $\sqrt{s}=130,200$~AGeV  
and $\rho_g (\tau_0) \sim 100  -  400/{\rm fm}^3$
in central $Pb+Pb$ reactions at $\sqrt{s}=5500$~AGeV,  since widely different 
models (e.g. see \cite{Wang:1991ht,Eskola:1999fc}) seem to 
be roughly consistent with data~\cite{Back:2000gw}. It is, therefore, 
essential to check the energy dependence of the density of the produced 
quark-gluon plasma (QGP) with observables complementary to  the  particle  
multiplicity $dN^{ch}/dy$ and  transverse energy  $dE_T/dy$ per unit rapidity. 
High-$p_T$ observables are  ideally suited for this task because they provide 
an estimate~\cite{Wang:xy} of the energy loss,  $\Delta E$,  of fast partons, 
resulting from medium induced non-Abelian radiation along their 
path, as first discussed in~\cite{Gyulassy:1993hr,Wang:1994fx} in the
context of relativistic heavy ion reactions. The approximate linear dependence 
of  $\Delta E$  on  $\rho_g$  is the key that enables high-$p_T$ observables  
to convey information 
about the  initial conditions. However, 
$\Delta E$ also depends non-linearly on the size, $L$, 
of the medium~\cite{Baier:1998n,Gyulassy:2000fs}   
and therefore differential observables which have well 
controlled geometric dependences are also highly desirable.

A new way to probe $\Delta E$ in variable geometries was recently  proposed  
in Refs.~\cite{Wang:2000fq,Gyulassy:2000gk}. The idea is to exploit the  
spatial  azimuthal asymmetry  of non-central  nuclear collisions. The 
dependence of $\Delta E$ on the path length $L(\phi)$ 
naturally results in a pattern of azimuthal asymmetry of high-$p_T$ hadrons 
which  can be measured  via the differential elliptic 
flow parameter (second Fourier coefficient),  
$v_2(p_T)$~\cite{Voloshin:1996,Ollitrault:bk}. 
Before we show the sensitivity of the high-$p_T$ $v_2(p_T > 2\;{\rm GeV})$ to 
different initial conditions we briefly discuss the various model 
calculations  for the ``elliptic flow'' coefficient $v_2$: 
\begin{enumerate}
\item The elliptic flow parameter $v_2$ was first introduced in the context of
{\em relativistic hydrodynamics}~\cite{Ollitrault:bk} and reflects the fact 
that due to the macroscopic sizes of large nuclei many aspects of $A+A$ 
collisions are  driven by nuclear geometry. In non-central 
collisions the interaction region has a  characteristic ``almond-shaped'' 
form as shown in Fig.~\ref{fig:ellips}. 
\vskip 0.7 cm
\begin{figure}[htb]
\begin{center}
\resizebox{125mm}{95mm}
{\includegraphics{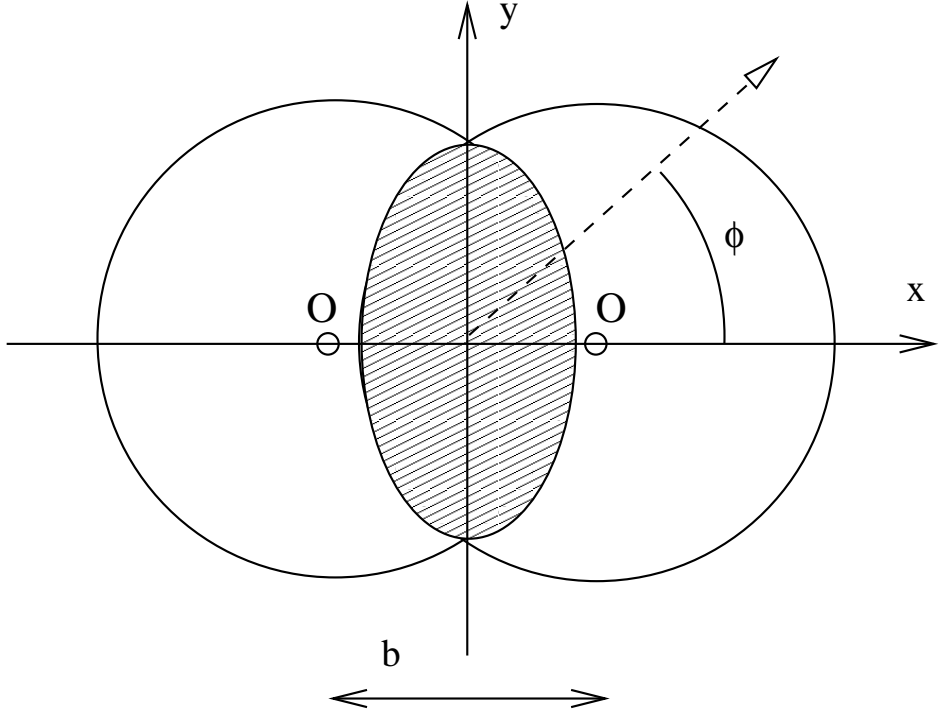}}
\vskip .2cm
\caption{\small The nuclear overlap region in non-central $A+A$ collisions
shows the importance of reaction geometry. Model calculation described here 
convert  the spatial anisotropy illustrated above into momentum 
anisotropy of  measured hadrons.}
\label{fig:ellips}
\end{center}
\end{figure}
Hydro calculations convert the ellipticity  of the reaction volume into 
momentum space azimuthal asymmetry  
\begin{equation}
\varepsilon=\frac{\langle x^2 \rangle - \langle y^2 \rangle}
{\langle x^2 \rangle + \langle y^2 \rangle }  \; \Longleftrightarrow \; 
\frac{\langle p_x^2 \rangle - \langle p_y^2 \rangle}
{\langle p_x^2 \rangle + \langle p_y^2 \rangle }  = 
\langle \cos 2 \phi \rangle =  
\frac{\int_{0}^{2 \pi}  d\phi\,\cos 2\phi\,
\frac{dN^h}{ dy\,p_{ T}\,dp_{T}\, d\phi }  }   
{ \int_{0}^{2 \pi}  d\phi\, \frac{dN^h}{ dy\,p_{ T}\,dp_{ T}\, d\phi  } }  
\; \; 
\label{anisrelat}  
\end{equation}
through the higher pressure gradient along the small axis. 
The elliptic flow is thus perfectly correlated to the 
reaction plane and can be used for its determination~\cite{Ollitrault:ba}. 
Hydrodynamic simulations~\cite{Kolb:2001qz,Teaney:2001av} typically describe 
well data from relativistic nucleus-nucleus collisions at 
$\sqrt{s}_{NN}=200$~GeV 
up to $p_T \simeq 1.5-2$~GeV and it is not unlikely that at LHC energies of  
$\sqrt{s}_{NN}=5.5$~TeV the region of validity of those calculations may extend
to $p_T \simeq 5$~GeV. 
 
\item  Initial conditions can also be mapped onto final state observable 
distributions  by solving covariant Boltzmann transport equations as in 
{\em cascade models} (partonic, hadronic, and multi-phase). Elliptic  flow in 
this approach is generated  via  multiple elastic scatterings. Calculations are 
sensitive to the choice of initial conditions~\cite{Molnar:2001ux} 
and are currently 
limited by statistics to $p_T \sim 6$~GeV. It is interesting to note that they 
can match the high-$p_T$ behaviour of the $v_2$ but require extremely large 
initial gluon rapidity densities $dN^g/dy \simeq 16000$~\cite{Molnar:2001ux} 
and/or string melting~\cite{Lin:2001zk}.
    
\item  Memory of the initial parton density, reaction geometry, and 
the consequent dynamical evolution is  also retained by large transverse 
momentum partons (and fragmented hadrons) 
through their {\em jet quenching} pattern~\cite{Wang:2000fq,Gyulassy:2000gk}.  
While this approach is discussed in more detail below, it is important to 
emphasize here that at the single inclusive jet (or hadron) 
level the resulting 
high-$p_T$ azimuthal asymmetry is also perfectly coupled to the 
reaction plane.  
It has been suggested that in the limit of very large energy loss the momentum
asymmetry is driven by jet production from the boundary of the interaction 
volume~\cite{Shuryak:2001me}.   

\item Recently, a classical computation of the elliptic flow  at transverse 
momenta $k_T^2 > Q_s^2$ in the  framework of {\em gluon saturation}  models 
has been performed~\cite{Teaney:2002kn}. It was found 
that the azimuthal asymmetry is generated already at proper 
time $\tau=0$, i.e. it is built in the coherent initial conditions. 
The resulting elliptic flow  coefficient was found to vanish quickly 
$v_2(k_T) \propto k_T^{-2} (R_x^{-2} - R_y^{-2})$ above $Q_s$ 
($\sim 1$~GeV for RHIC  and $\sim 1.4-2$ for LHC energies) which is  not 
supported by the current data.     

\item An approach that {\em does not} associate azimuthal asymmetry with the 
reaction plane has also been presented~\cite{Kovchegov:2002nf}. Both  
high-$p_T$ and low-$p_T$ $v_2$ emerge as a {\em back-to-back jet correlation 
bias}  (with arbitrary direction relative to the reaction geometry for every 
$p_T$ bin). For large transverse momenta $v_2 \propto \ln p_T/\mu$ 
suggest an easily detectable factor of 3 increase in going 
from  $p_T=5$~GeV to $p_T = 100$~GeV at LHC. The $p_T$-integrated 
$v_2 \propto 1/Q_s$ at LHC exhibits $\sim 50\%$ reduction relative to RHIC. 
(It can also be deduced  that $v_2$ is larger at the SPS in comparison 
to RHIC in this model.)   

\end{enumerate}

The methods for $v_2$ analysis can be broadly divided in two  categories:
two-particle methods discussed, e.g., in~\cite{Wang:qh} and multi-particle 
methods~\cite{Borghini:2001vi,Borghini:2001zr}. 
In two-particle methods the error on the determined $v_2$ from non-flow 
(non-geometric) correlations is ${\cal O}(1/(v_2 M))$, where $M$ is the 
measured multiplicity.  With multi-particle methods this error goes down 
typically to ${\cal O} ( 1/(v_2 M^2))$, i.e., smaller by a factor of order 
$M$. Although it is not possible to {\em completely} eliminate the non-flow 
components to $v_2$, experimental techniques based on higher oder cumulant 
analysis~\cite{Borghini:2001vi,Borghini:2001zr} will be able in many 
cases to  {\em clearly distinguish} between between reaction geometry 
generated  azimuthal asymmetry and back-to-back jet bias.

\section{\bf Energy loss in a longitudinally expanding plasma}

We first generalize the finite energy gluon bremsstrahlung 
theory~\cite{Gyulassy:2000fs,Gyulassy:1999ig,Gyulassy:1999zd,Gyulassy:2000er} 
to take into account the  expansion  (neglected in~\cite{Wang:2000fq}) of 
the produced gluon-dominated  plasma  while retaining kinematic  constraints 
important for  intermediate jet energies. 
The GLV reaction operator formalism~\cite{Gyulassy:2000er} expands
the radiative energy loss formally in powers of the mean 
number, $\chi$, of interactions that the jet+gluon system
suffer along their path of propagation through dense matter.
For a jet produced at point $\vec{x}_0$, at time $\tau_0$,
in an {\em expanding}  and possibly azimuthally asymmetric
gluon plasma of density $\rho(\vec{x},\tau)$,
the opacity in direction $\hat{v}(\phi)$ is 
\begin{equation}
\chi(\phi) =\int_{\tau_0}^\infty d\tau \; 
\sigma(\tau)\rho(\vec{x}_0 + \hat{v}(\phi)(\tau-\tau_0),\tau)\;.
\label{opac}
\end{equation}
Note that  the gluon-gluon elastic cross section,  
$\sigma(\tau)=9 \pi \alpha_s^2/2\mu^2_{eff}(\tau)$,  and
the density may  vary along the path.  
The explicit closed form expression for the $n^{\rm th}$ order
opacity expansion of the gluon radiation double differential distribution
for a static medium is given in~\cite{Gyulassy:2000er}. 
The generalization to the case of arbitrary medium similar to 
Eq.~(\ref{opac}) is trivial and  amounts to replacing the static weight 
with the nested position integrals in the case of a dynamically evolving 
quark-gluon  plasma: 
\begin{equation}
\frac{1}{n!} \left[\frac{L}{\lambda_g}\right]^n  \;\; \Longrightarrow
\;\; \prod\limits_{i=1}^n \, \int_0^{L-\tau_1 - \cdots - \tau_{i-1} }
\frac{d \tau_i}{\lambda_g(\tau_i)} \;. 
\end{equation}
Fortunately, the opacity expansion converges very rapidly, and the first 
order term  was found  to give the dominant 
contribution. Higher order corrections decrease rapidly with  the jet 
energy $E$. All numerical results discussed here include  2$^{\rm nd}$ and 
3$^{\rm rd}$ order correlations between scattering centers that can alter 
the leading order dependence outlined analytically below. 

In the case of 1+1D Bjorken longitudinal expansion with initial plasma 
density $\rho_0=\rho(\tau_0)$ and formation time  $\tau_0$, i.e.   
\begin{equation} 
\rho(\tau)=\rho_0\left(\frac{\tau_0}{\tau}\right)^\alpha \;, 
\label{bjork}
\end{equation} 
it is possible to obtain a closed form analytic formula~\cite{Gyulassy:2000gk}
(under the strong {\em no kinematic bounds} assumption) 
for the energy loss due to the dominant first order 
term~\cite{Gyulassy:2000fs}. 
For a hard jet penetrating the quark-gluon plasma 
${d \Delta E^{(1)}}/{E \, dx} \propto  
\int^\infty_{\tau_0}  f(Z(x,\tau)) \,  {d \tau}/{\lambda(\tau)}$, 
where $x \simeq \omega/E$ is the momentum fraction of the 
radiated gluon and   the formation physics function $f(Z(x,\tau))$ is 
defined in~\cite{Gyulassy:2000gk} to be 
\begin{equation}
f(x,\tau)= \int_0^\infty \frac{du}{u(1+u)}   
\left[ 1 -   \cos \left (\,u Z(x,\tau) \right) \right]  \;. 
\label{fz} 
\end{equation}
With $Z(x,\tau)=(\tau-\tau_0)\mu^2(\tau)/2 x E$ being the local 
formation physics  parameter,  two simple analytic limits of  
Eq.~(\ref{fz}) can be obtained. 
For $x \gg x_c =  \mu(\tau_0)^2 \tau_0^{ \frac{2\alpha}{3}}  
L^{1- \frac{2\alpha}{3} }/2E  = L\mu^2(L)/2 E$, in which case  
the formation length is large compared to the size of the medium,  
the small $Z(x,\tau)$ limit applies leading to  $f(Z) \approx \pi Z / 2 $. 
The interference pattern along the  gluon path becomes important and accounts 
for the the non-trivial dependence of the energy loss on $L$.   
When $x \ll x_c$, i.e. the formation length  is small compared to the plasma 
thickness,  one gets $f(Z) \approx \ln Z$.  In the $x \gg x_c$ 
limit~\cite{Gyulassy:2000gk} 
the radiative spectrum reads:
\begin{equation}
\frac{d \Delta E^{(1)}_{x\gg x_c}}{dx} \approx 
\frac{C_R \alpha_s}{2(2-\alpha)}  
\frac{\mu(\tau_0)^2\tau_0^\alpha 
L^{2-\alpha}}{\lambda(\tau_0)}  \frac{1}{x} \;.  
\label{dIdxxbig}  
\end{equation}
The mean  energy loss (to first order in $\chi$) is given by 
\begin{equation}
\Delta E^{(1)} =  \frac{C_R \alpha_s}{2(2-\alpha)}  
\frac{\mu(\tau_0)^2\tau_0^\alpha L^{2-\alpha} }{\lambda(\tau_0)}  
\left( \ln \frac{2 E}{\mu(\tau_0)^2 \tau_0^{ \frac{2\alpha}{3}}  
L^{1- \frac{2\alpha}{3} }}    +  \cdots \right)  \;.   
\label{totde} 
\end{equation}
The logarithmic enhancement with energy comes from the  
$x_c< x <1$ region~\cite{Gyulassy:2000er}. 
In the case of sufficiently large jet energies  
($E\rightarrow \infty$) this term dominates. For parton energies   
$<20$~GeV, however, corrections to this leading $\ln 1/x_c$  expression 
that can be exactly evaluated numerically from the GLV expression 
and are found to be  comparable in size. The  effective $\Delta E/E$ in this 
energy range was found to have a plateau~\cite{Levai:2001hd}. 
We note that for Bjorken expansion with $\alpha=1$, the asymptotic 
energy loss can be expressed
in terms of the initial gluon rapidity density as
\begin{equation}
\Delta E_{\alpha=1}(L) \approx \
\frac{9C_R\pi\alpha_s^3}{4} \left(\frac{1}{\pi R^2} 
\frac{dN^g}{dy} \right) \, L\,\ln \frac{1}{x_c} \;. 
\label{debjj}
\end{equation}
It is evident from Eq.~(\ref{debjj}) that the appropriate variable to drive 
non-Abelian jet energy loss calculations with is the effective initial 
gluon rapidity density $dN^g/dy$. 

\section{Parton energy loss and  nuclear geometry}

For nucleus-nucleus collisions  the co-moving  plasma  produced in an $A+B$ 
reaction at impact parameter $b$ at formation time 
$\tau_0$ has a transverse coordinate distribution at mid-rapidity  
$\rho_g({\bf r},z=0,\tau_0)$.  In studying jet production and 
propagation in nuclear environment it is not always  technically possible 
to perform the Monte-Carlo averaging over the 
jet production points coincidentally with the simulation of 
parton fragmentation.  It is therefore useful to separate the medium 
dependence of the mean jet  energy loss as a function of the extent 
of the nuclear matter traversed and the
azimuthal angle $\phi$ relative to the reaction plane. 
In the Bjorken, Eq.~(\ref{bjork}), and linear $f(Z)\approx \pi Z/2$, 
Eq.~(\ref{fz}), approximations  the total energy loss is proportional to 
a line  integral along the jet trajectory
${\bf r}(\tau,\phi)={\bf r}+\hat{v}(\phi)(\tau-\tau_0)$,
averaged over the distribution  of the jet production points 
\begin{eqnarray}
F(b,\phi) &=&  \int d^2{\bf r} \; \frac{T_A(r)T_B(|{\bf r}-{\bf b}|)}
{T_{AB}(b)}  \int_{\tau_0}^\infty
 d \tau \; \tau\; \left(\frac{\tau_0}{\tau}\right)^\alpha 
\rho_0({\bf r}+\hat{v}(\phi)(\tau-\tau_0))\;.
\label{linint}
\end{eqnarray}

\vskip 0.5cm
\begin{figure}[htb]
\begin{center} 
\hspace*{-.4in}\epsfig{file=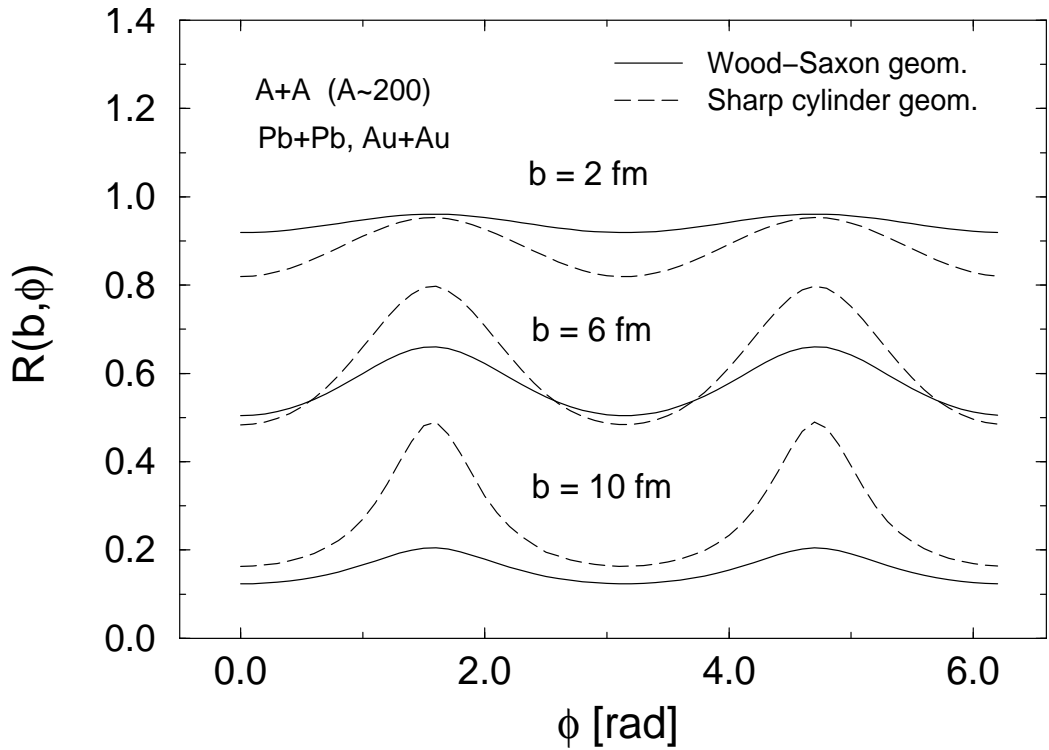,height=6.in,width=4.5in,
bbllx=90,bblly=0,bburx=600,bbury=700, clip=,angle=-90}
\vspace*{-.4in}
\caption{\small The modulation function  $R(b,\phi)$   
is plotted versus $\phi$ for several impact parameters $b=2, 6, 10$~fm
from Ref.~\protect{\cite{Gyulassy:2000gk}}.
Diffuse Wood-Saxon versus uniform sharp 
cylinder geometries are compared. 
The most drastic difference between these geometries
occurs at  high impact parameters.}
\label{ng:fig2-vv}
\end{center} 
\end{figure}
\noindent 
$T_A(r)=\int dz\, \rho_A({\bf r},z)$ and  $T_{AB}(b) = \int d^2{\bf r}  \, 
T_A({\bf r})T_B({\bf r- b})$ depend on the geometry.
In particular, for a sharp uniform cylinder of radius $R_{\rm eff}$
 one readily gets ${T_A}(r)=(A/\pi R^2_{\rm eff})
\theta (R_{\rm eff}-|{\bf r}|)$ and ${T_{AB}}(0)=A^2/\pi R^2_{\rm eff}$.
We can therefore define the  effective radius of the sharp cylinder 
equivalent to a diffuse Wood-Saxon geometry via
$ F (0,\phi)_{\rm Wood-Saxon} =  F (0,\phi)_{\rm Sharp \;\; cylinder}$.
For $Au+Au$ collisions and $\alpha=1$
the above constraint gives $R_{\rm eff}\approx 6$~fm. 

For a non-vanishing  impact parameter $b$ 
and jet direction ${\hat{v}(\phi)}$,  we calculate the energy loss as
\begin{equation}
\frac{\Delta E(b,\phi )}{E}
= \frac{F(b,\phi )}{F(0,\phi)} \, \frac{\Delta E(0)}{E} 
\equiv R(b,\phi ) \, \frac{\Delta E(0)}{E} \;,
\label{separat}
\end{equation}
where the modulation function $ R(b,\phi)$ captures in the 
{\em linearized} approximation the $b$ and $\phi$ 
dependence of the jet energy loss and also provides a rough estimate 
of the maximum ellipticity generated via correlations to the 
reaction plane.
Fig.~\ref{ng:fig2-vv} shows the  $R(b,\phi)$  modulation factor
plotted against the azimuthal angle $\phi$ for impact parameters 
$b=2, 6, 10$~fm. Note that $R(b,\phi)$ reflects not only 
the dimensions of the characteristic ``almond-shaped'' cross section 
of the interaction volume but also the rapidly decreasing  initial plasma 
density  as a function of the impact parameter.

In order to compare to data at $p_T < 2$~GeV at RHIC and  $p_T < 5$~GeV  
at LHC, one must also take into account the  soft non-perturbative component
that cannot be computed with the eikonal jet quenching formalism.
The hydrodynamic elliptic flow~\cite{Ollitrault:bk} was found 
in~\cite{Kolb:2001qz} to have the monotonically growing form
$v_{2s}(p_T) \approx {\rm tanh}(p_T/(10\pm 2\;{\rm GeV}))$ 
at $\sqrt{s}=200$~AGeV and to be less sensitive to the initial conditions
than the high-$p_T$  jet quenching   studied  here. The interpolation between 
the low-$p_T$ relativistic hydrodynamics region and the high-$p_T$ 
pQCD-computable region can be evaluated as in~\cite{Gyulassy:2000gk}. 
 
\vskip 0.5cm
\begin{figure}[htb]
\begin{center} 
\hspace*{-.4in}\epsfig{file=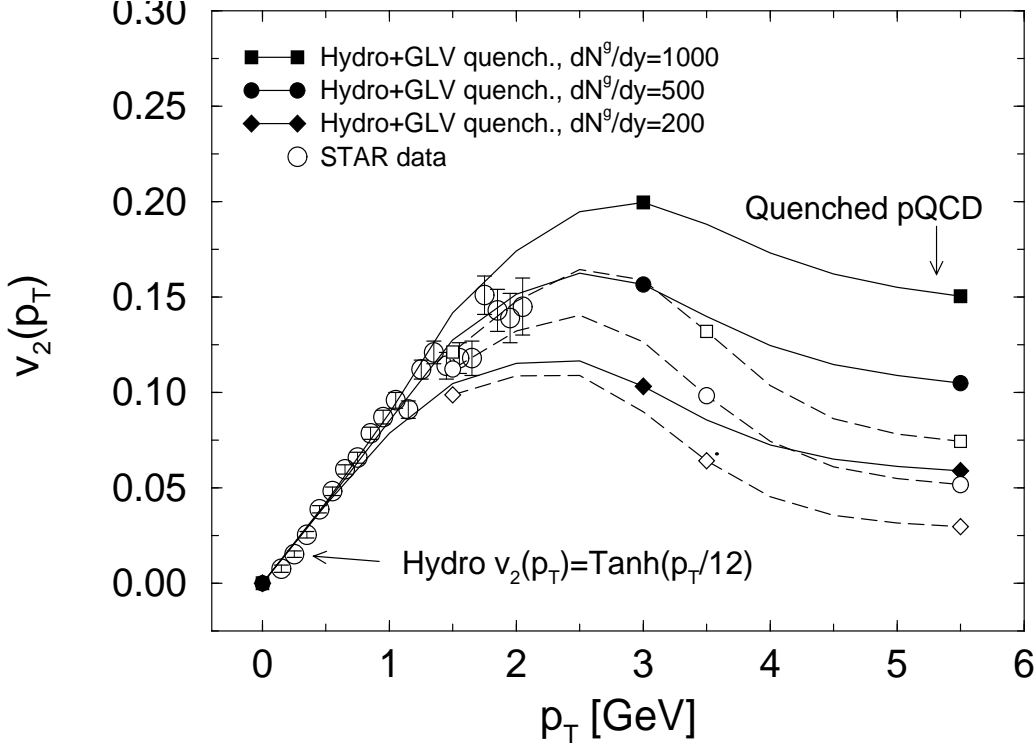,height=6.in,width=4.5in,
bbllx=90,bblly=0,bburx=600,bbury=700, clip=,angle=-90}
\vspace*{-.4in}
\caption{\small The interpolation of $v_2(p_T)$
between the soft hydrodynamic~\protect{\cite{Kolb:2001qz}} 
and hard pQCD regimes is shown for $b=7$ fm adapted from 
Ref.~\protect{\cite{Gyulassy:2000gk}}.
Solid (dashed) curves  correspond to sharp cylindrical (diffuse 
Wood-Saxon) geometries presented in Fig.~\ref{ng:fig2-vv}.}
\label{ng:fig4-vv}
\end{center} 
\end{figure}

Fig.~\ref{ng:fig4-vv} shows the predicted pattern of high-$p_T$ anisotropy.
Note the difference between sharp cylinder and diffuse Wood-Saxon geometries
at $b=7$~fm  approximating roughly  20-30\% central events.  
While  the central ($b = 0$)  inclusive quenching 
is insensitive to the density profile, non-central events clearly exhibit  
large sensitivity to the actual distribution. 
We conclude that $v_2(p_T>2\;{\rm GeV},b)$ 
provides essential complementary information about the geometry and impact
parameter dependence of the initial conditions in $A+A$. In particular, the 
rate at which the $v_2$  coefficient  decreases at high $p_T$ is an  
indicator of the diffuseness of that geometry. Minimum bias STAR data at
RHIC~\cite{Filimonov:2002xk,Jacobs:2002pz} for $p_T \geq 6$~GeV  now seem to 
support the predicted~\cite{Wang:2000fq,Gyulassy:2000gk} 
slow decrease of $v_2$ at large transverse momenta.      
Recently in~\cite{Vitev:2002pf} hadron suppression in $Au+Au$ ($Pb+Pb$) 
relative to the binary scaled $p+p$ result at $p_T \simeq 5$~GeV for RHIC 
conditions ($\sqrt{s}_{NN}, dN^g/dy$) was found to be approximately equal 
to the quenching factor at LHC at a much larger transverse momentum scale 
$p_T\simeq 50$~GeV. One may thus anticipate proportionally large 
($\sim 10-15\%$) azimuthal asymmetry for high $p_T$ at the LHC. 

\section{\bf Energy loss of jets in transversely expanding medium} 
  
Transverse expansion can only be very crudely modelled in Eq.~(\ref{bjork}) by
taking $\alpha > 1$. Such approach neglects the time needed for the rarefaction 
wave to propagate from the surface to the center of the interaction region.
To derive an improved  analytic expression taking transverse flow into account, 
we consider next an asymmetric expanding sharp {\em elliptic}  
density profile the surface of which is defined by  
\begin{equation}  
\frac{x^2}{(R_x+v_x \tau)^2} +\frac{y^2}{(R_y+v_y \tau)^2} = 1 \;. 
\label{ellip}  
\end{equation}  
The area of this elliptic transverse profile increases with time, $\tau$,  
as  $A_\perp(\tau)=\pi (R_x+v_x \tau) (R_y+v_y \tau)$. 
The plasma density seen by a jet in direction  
$\phi_0$ starting from ${\bf x}_0=(x_0,y_0)$ inside the ellipse  
with a specified initial condition 
$\tau_0\rho_0= 1 / (\pi R_xR_y)\, dN^g/dy$ is   
\begin{eqnarray}  
\rho(\tau,\phi_0;x_0,y_0)&=&   \frac{1}{\pi}  
\frac{dN^g}{dy}\left(\frac{1}{\tau}\right) \left(\frac{1}{R_x+v_x \tau}\right)  
 \left(\frac{1}{R_y+v_y \tau}\right)  
 \nonumber \\[1ex]  
&\;& \times \theta\left(1-\frac{(x_0+\tau \cos \phi_0)^2}{(R_x+v_x \tau)^2}   
+\frac{(y_0+\tau \sin \phi_0)^2}{(R_y+v_y \tau)^2}\right) \;.  
\label{rhoex}  
\end{eqnarray}
We approximate the assumed $\phi_0$ independent screening  
$\mu(\tau)\approx g T(\tau) = 2 (\rho(\tau)/2)^{1/3}$ since  
$g \simeq 2$ and $\rho = (16 \zeta(3)/ \pi^2) T^3 \simeq 2 T^3 $ for  
gluon plasma  and define   
$\tau(\phi_0)$ as the  escape time to reach the expanding elliptic surface from  
an initial point ${\bf x}_0$ in the azimuthal direction $\phi_0$.  
We take $\omega(\phi_0)=2\, \tau(\phi_0) \,( \rho(\tau(\phi_0))/2 )^{2/3}$  
to estimate an upper bound on the logarithmic enhancement factor.  
A short calculation leads to: 
\begin{eqnarray}
\Delta E^{(1)}(\phi_0)   
  &\approx&  \frac{9}{4} \, \frac{ C_R\alpha_s^3}{R_x R_y}\frac{dN^g}{dy}  
\; \frac{\ln   \frac{1 + a_x \tau(\phi_0)}{1+ a_y \tau(\phi_0)}}{a_x-a_y}  
\; \ln \frac{E}{\omega(\phi_0)}  \;\; , 
\label{deaz}   
\end{eqnarray}
where $a_x=v_x/R_x, a_y=v_y/R_y$. 
This expression is a central result for transversely expanding media
and provides a simple analytic  generalization 
that interpolates between pure Bjorken 1+1D expansion for small 
$a_{x,y} \tau$, and 3+1D expansion at large $a_{x,y} \tau$. 
 
In the special case of pure Bjorken (longitudinal) expansion 
with $v_x=v_y=0$  
\begin{eqnarray}
\Delta E^{(1)}_{Bj}(\phi_0)&=&   
\frac{9C_R\alpha_s^3}{4 R_x R_y}   
\frac{dN^g}{dy}  \, \tau(\phi) \,\ln \frac{E}{\omega(\phi_0)}  \; . 
\label{debj} 
\end{eqnarray}
In this case, the energy loss depends  {\em linearly} on  $\tau(\phi)$ 
and we recover the result of Eq.~(\ref{debjj}). 
Another special case is azimuthally {\em isotropic} expansion 
with $R_x=R_y=R$ and $v_x=v_y=v_T$. Taking also  the longitudinal 
Bjorken expansion into account leads in this case to 
\begin{eqnarray}  
\Delta E^{(1)}_{3D}(\phi_0)&=&   
\frac{9}{4} \frac{C_R\alpha_s^3}{R^2}   
\frac{dN^g}{dy}  \, \frac{\tau(\phi_0)}{1+v_T\tau(\phi_0)/R}  
\,\ln \frac{E}{\omega(\phi_0)} \; .   
\label{debj1}  
\end{eqnarray}
We note that for a jet originating near the center of the medium  and  
{\em fully penetrating} the plasma the enhanced escape time  
due to expansion $\tau=R/(1-v_T)$ compensates for the 
$1/(1+v_T\tau(\phi_0)/R)$ dilution factor. Therefore, in this isotropic case, 
the extra dilution due to transverse expansion 
 has in fact no effect of the total energy loss: 
\begin{equation}
\Delta E^{(1)}_{1D}(b=0\; {\rm fm}) 
\approx \Delta E^{(1)}_{3D}( b=0\; {\rm fm}) \;, 
\label{aprind}
\end{equation}
modulo logarithmic factors which become sizable  only for large $v_T$.  
An important consequence is that the  inclusive azimuthally averaged jet quenching
pattern in central  collisions is approximately independent of transverse 
expansion.  We have checked numerically that  Eq.~(\ref{aprind}) holds
for realistic transverse density profiles~\cite{Kolb:2001qz}. 

In non-central collisions, the azimuthal asymmetry of the  mean energy loss can be 
expanded in a Fourier series and characterized as 
\begin{eqnarray}
\Delta E^{(1)}_{3D}(\phi)&=&\Delta E(1+ 2 \delta_2(E)\cos 2\phi +\cdots) \; .   
\label{debjv2}  
\end{eqnarray}
It is correlated to the final measured elliptic ``flow'' of jets and hadrons 
and  has been evaluated by using a full
hydrodynamic calculation from  Ref.~\cite{Kolb:2001qz}. In this case we use the
parameterization eBC of~\cite{Kolb:2001qz} to initialize the system and treat
gluon number as conserved current to calculate the density evolution needed
in the line integral Eq.~(\ref{linint}), where it replaces the naive 
Bjorken $(\tau_0/\tau)^\alpha$ expansion. We average over the jet
formation points the density of which is given by
the number of binary collisions per unit area as in the Woods-Saxon geometry
used in  Ref.~\cite{Gyulassy:2000gk}. We find that the  azimuthal asymmetry of 
the energy loss  is strongly  reduced for realistic hydrodynamic flow 
velocities. This implies a much smaller $v_2$ at high $p_T$  than obtained in 
Ref.~\cite{Gyulassy:2000gk} where transverse  
expansion was not considered and poses questions about the observability of the
effect at LHC. 

\section{LHC-specific remarks}
\label{sbsec:remarks}

There are several important aspects in which LHC and RHIC will differ 
significantly. We briefly discuss the implications of those differences 
for high-$p_T$ $v_2$ measurements:   

\begin{enumerate}

\item    Currently at RHIC at $\sqrt{s}= 200$~AGeV the $p_T \geq 2-3$~GeV 
regime is perturbatively  computable~\cite{Vitev:2002pf} (modulo 
uncertainties in the baryonic sector~\cite{Vitev:2001zn}).  
At LHC the $p_T$ region which is not accessible through the 
pQCD approach may extend to transverse momenta as high as 5-10~GeV. This
would imply the validity of the relativistic hydrodynamics in this
domain, the extent of which  can be tested by looking 
for marked deviations in the growth of $v_2(p_T)$, saturation, and turnover.
 
\item   Estimates of the initial gluon rapidity density at LHC vary from 
$dN^g/dy=2500$  to  $dN^g/dy=8000$. This would imply very large parton energy 
loss, at least in some regions of phase space. In this case jet production 
for moderate transverse momenta may be limited to a small shell on 
the surface of the interaction region,  leading to a constant $v_2(p_T)$ 
purely determined by geometry~\cite{Shuryak:2001me}. 

\item    Since mean transverse expansion velocities at RHIC have been  
estimated to be on the order of $v_T \simeq 0.5$ 
through relativistic hydrodynamics fits,
it is natural to expect even larger  values at LHC. This may lead to a 
significant reduction of the observed azimuthal asymmetry as discussed above. 
An important prediction of the approach put forth in~\cite{Gyulassy:2000gk} is 
that  $v_2(p_T)$ exhibits a slow decrease with increasing transverse momentum. 
This can be  used to distinguish  azimuthal anisotropy generated 
through energy loss  from alternative mechanisms.  

\end{enumerate}

\subsection{\bf Jet impact parameter dependence at the LHC} 

In light of the discussion in Sec.~\ref{sbsec:remarks}  it is important to 
asses the feasibility of  azimuthal asymmetry measurements for 
large-$E_T$ jets via  detailed simulations.
The impact parameter dependence of jet rates in $Pb+Pb$ collisions at the 
LHC was analyzed in~\cite{Lokhtin:2000}. The initial jet spectra at 
$\sqrt{s} = 5.5$~TeV were generated with PYTHIA$5.7$~\cite{pythia}. 
The initial distribution of jet pairs over impact
parameter $b$ of $A+A$ collisions (without  collective nuclear effects) was 
obtained by multiplying the corresponding nucleon-nucleon jet cross section, 
$\sigma _{NN}^{\rm jet}$, by the number of binary 
nucleon-nucleon sub-collisions~\cite{Vogt:1999}: 
\begin{equation} 
\label{jet_prob}
\frac{d^2 \sigma^0_{\rm jet}}{d^2b}(b,\sqrt{s})=T_{AA} (b)
\sigma _{NN}^{\rm jet} (\sqrt{s})
\left[ 1 - \left( 1- \frac{1}  
{A^2}T_{AA}(b) \sigma^{\rm in}_{NN} (\sqrt{s}) \right) ^{A^2} \right]   
\end{equation} 
with the total inelastic non-diffractive nucleon-nucleon cross section  
$\sigma^{\rm in}_{NN} \simeq 60$ mb. 

The rescattering and energy loss of jets in a gluon-dominated plasma, created 
initially in the nuclear overlap zone in $Pb+Pb$ collisions at different 
impact  parameters, were simulated. For details of this model one 
can refer to~\cite{Lokhtin:2000,Lokhtin:2001}. To be specific, we treated 
the medium as a boost-invariant longitudinally expanding fluid according to 
Bjorken's solution~\cite{Bjorken:1983} and used the initial conditions 
expected for central $Pb+Pb$ collisions at
LHC~\cite{Eskola:1994,Eskola:1997,Eskola:1998}: 
formation time $\tau_0 \simeq 0.1$ fm/c, initial temperature 
$T_0 \simeq 1$ GeV, gluon plasma density $\rho_g \approx 1.95T^3$. For our 
calculations we have used the collisional part of the energy loss and 
the differential scattering  cross section from~\cite{Lokhtin:2000}; 
the energy spectrum of coherent  medium-induced gluon radiation was 
estimated using the BDMS  formalism~\cite{Baier:1998n}. 

\begin{figure}[hbtp] 
\begin{center} 
\resizebox{105mm}{105mm} 
{\includegraphics{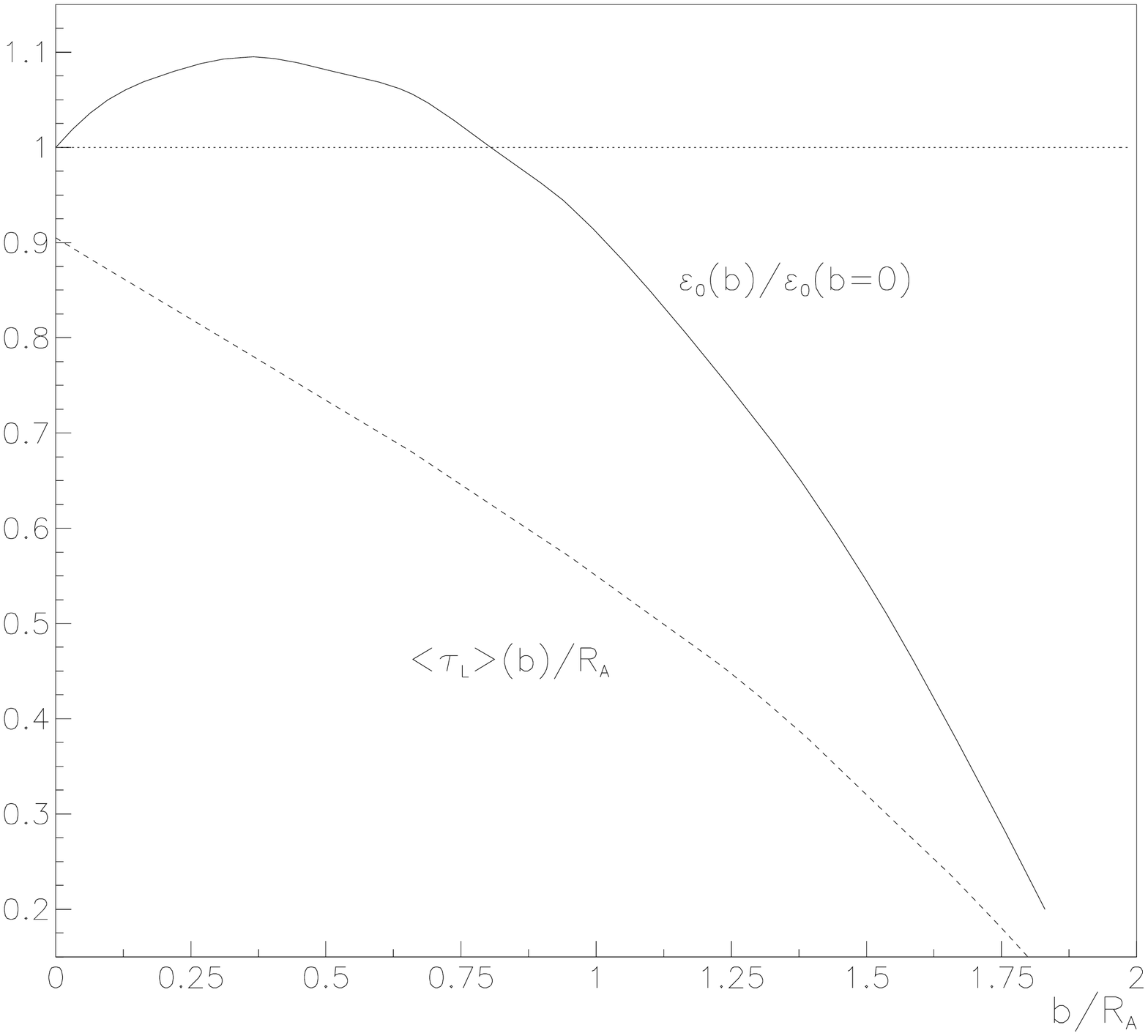}} 
\vskip -1.3 cm 
\caption{\small The impact parameter dependence of the initial energy density 
$\varepsilon_0 (b) / \varepsilon_0 (b=0)$ in nuclear overlap zone 
(solid curve), and  the average proper jet escape  time 
$\left< \tau_L \right> / R_A$ of from the dense matter 
(dashed curve)~\cite{Lokhtin:2000}.} 
\label{ng:fig3}

\vskip 0.7 cm 

\resizebox{110mm}{110mm} 
{\includegraphics{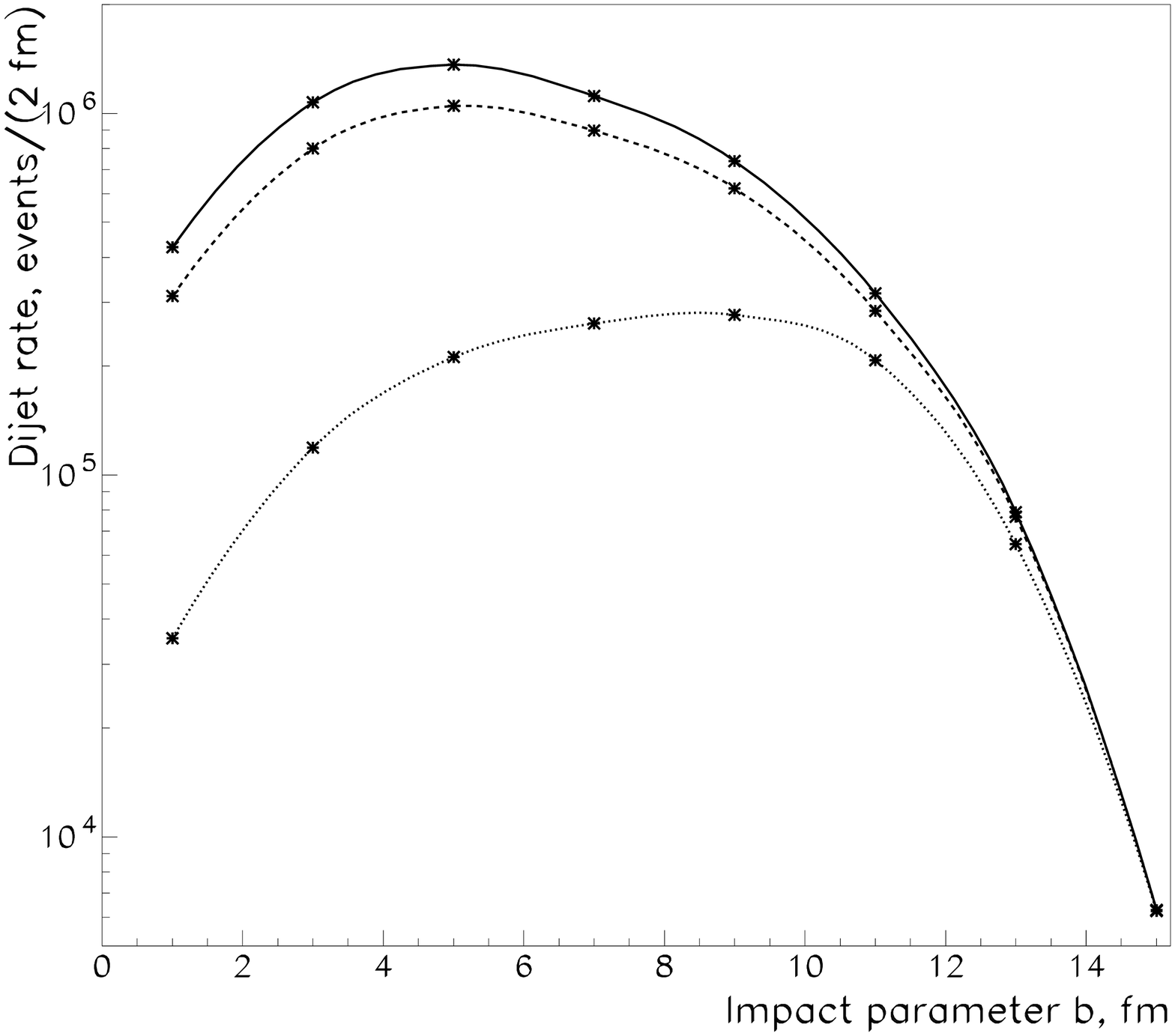}} 
\vskip -1.2cm 
\caption{\small The jet$+$jet rates for $E_T^{\rm jet} > 100$ GeV and 
$|\eta^{\rm jet}| < 2.5$ in 
different impact parameter bins: without energy loss (solid curve), with 
collisional loss (dashed curve), with collisional and radiative loss (dotted
curve)~\cite{Lokhtin:2000}.} 
\label{ng:fig4}
\end{center} 
\end{figure}

The impact parameter dependences of the initial energy density $\varepsilon_0$ 
and  the averaged over $\varphi$  jet escape  time $\left< \tau_L \right>$  
from the dense zone are shown in Fig.~\ref{ng:fig3}~\cite{Lokhtin:2000}.   
$\left< \tau_L \right>$  goes down almost linearly with increasing impact 
parameter $b$. On the other hand,  $\varepsilon_0$ is very weakly dependent 
of $b$ ($\delta \varepsilon_0 /\varepsilon_0  
 \mathrel{\lower.9ex \hbox{$\stackrel{\displaystyle  <}{\sim}$}}  10 \%$) 
up to $b$ on the order of nucleus radius $R_A$, and decreases rapidly only 
at $b  \mathrel{\lower.9ex \hbox{$\stackrel{\displaystyle  >}{\sim}$}}  R_A$. 
This suggests that for impact parameters $b < R_A$, 
where $\approx 60 \%$ of jet   pairs are produced, the difference in 
rescattering intensity and   energy loss is determined mainly by the 
different path lengths rather than the  initial energy density.  

Figure~\ref{ng:fig4} shows dijet rates in different impact parameter bins for 
$E_T^{\rm jet} > 100$ GeV and the pseudorapidity acceptance of central part 
of the CMS  calorimeters, $|\eta^{\rm jet}| < 2.5$, for three cases: 
$(i)$ without energy loss, $(ii)$ with collisional loss only, $(iii)$ with 
collisional and radiative loss. The total impact parameter integrated rates 
are normalized to the expected number of $Pb+Pb$ events during a two week LHC 
run,  $R= 1.2 \times 10^6$ s, assuming luminosity 
$L = 5 \times 10^{26}~$cm$^{-2}$s$^{-1}$. The maximum and mean values of 
$dN^{\rm dijet}/db$ distribution get shifted towards the larger $b$, because 
jet quenching is much stronger in central collisions than in peripheral ones. 
Since the coherent Landau-Pomeranchuk-Migdal radiation induces a strong 
dependence of the radiative energy loss of a jet on the angular cone 
size~\cite{Lokhtin:1998,Baier:1998}, the corresponding result for jets with 
non-zero cone size $\theta_0$ is expected to be somewhere between $(iii)$ 
($\theta_0 \rightarrow 0$) and $(ii)$ cases. Thus the observation of a 
dramatic change in the $b$-dependence of dijet rates in heavy ion collisions 
as compared  to what is expected from independent nucleon-nucleon reaction 
pattern, would indicate the existence of medium-induced parton rescattering. 

Of course, such kind of measurements require the adequate determination of 
impact parameter in nuclear collision with high enough accuracy. 
It has been shown in~\cite{Damgov:2001} that for the CMS experiment the 
very forward pseudorapidity region  $3 \le |\eta| \le 5$ can provide a 
measurement of impact parameter via  the  energy flow in the very forward (HF) 
CMS calorimeters with resolution  $\sigma_b \sim 0.5$ fm for central and 
semi-central $Pb+Pb$  collisions (see details in  the section on jet 
detection at CMS). 

\subsection{\bf Jet azimuthal anisotropy at the LHC} 

While at RHIC the {\em hadron}  azimuthal asymmetry at high-$p_T$  is 
being analyzed,  at LHC energies one can hope to observe  similar effects 
for the hadronic jet itself~\cite{Lokhtin:2001} due to the large inclusive 
cross section  for  hard jet  production  on  a  scale  $E_T \sim 100$ GeV. 
 
\begin{figure}[htb]
\begin{center} 
\resizebox{115mm}{115mm} 
{\includegraphics{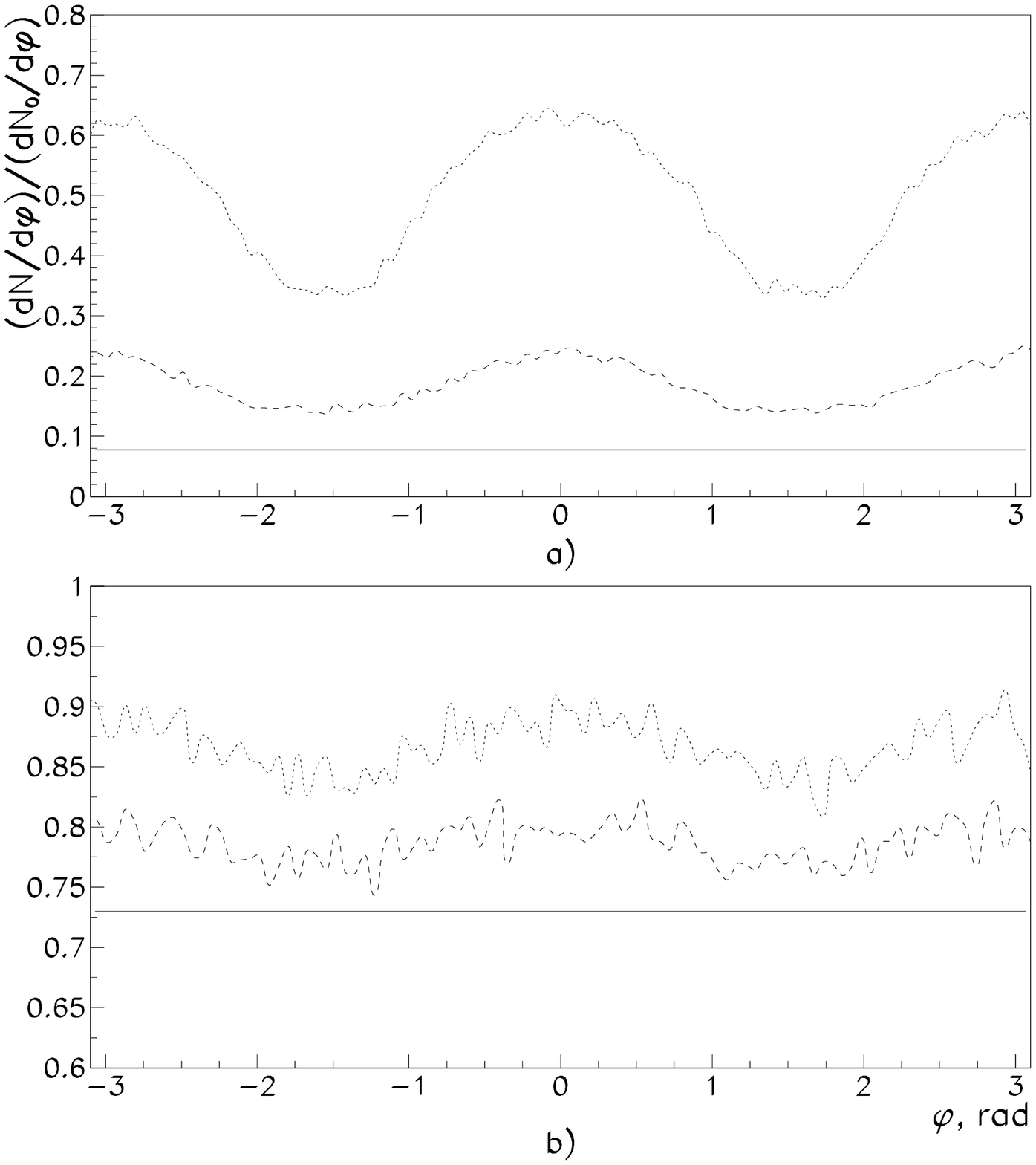}} 
\caption{\small The jet distribution over azimuthal angle for the cases with 
collisional and radiative loss (a) and collisional loss only (b), 
$E_T^{\rm jet} > 100$ GeV and $|\eta^{\rm jet}| < 
2.5$~\cite{Lokhtin:2001}. The histograms (from bottom to top) correspond 
to the impact parameter values $b = 0$, $6$ and $10$ fm.} 
\label{ng:fig5}
\end{center} 
\end{figure}

The  anisotropy of the energy loss ($\Delta E$)  goes up with increasing $b$, 
because the azimuthal  asymmetry of the  interaction volume gets stronger. 
However, the absolute  value of the energy loss goes down with increasing $b$ 
due to the reduced  path length $L$ (and $\varepsilon _0$ at 
$b \mathrel{\lower.9ex \hbox{$\stackrel{\displaystyle  >}{\sim}$}}  R_A$, 
see Fig.~\ref{ng:fig3}). The non-uniform dependence of $\Delta E$ on 
the azimuthal angle $\varphi$ is then mapped onto the jet spectra in  
semi-central collisions. Figure~\ref{ng:fig5} from~\cite{Lokhtin:2001} shows 
the distribution of jets over $\varphi$ for the cases 
with collisional and radiative  loss (a) and collisional loss only (b) for 
$b = 0$, $6$ and $10$ fm. The same conditions and kinematical acceptance as
in Fig.~\ref{ng:fig4} were fulfilled.  
The distributions are normalized by the distributions of jets as a function of
$\varphi$ in $Pb+Pb$ collisions without energy loss. 
The azimuthal anisotropy becomes stronger in going from central to semi-central
reactions, but the absolute suppression factor is reduced with increasing $b$. 
For jets with finite cone size one can expect the intermediate result between
cases (a) and (b), because, as we have mentioned before, radiative loss
dominates at relatively small angular sizes of the jet cone 
$\theta_0 (\rightarrow 0)$,  while the relative contribution of collisional 
loss grows with increasing $\theta_0$. 

In non-central collisions the jet distribution over $\varphi$ is approximated 
well by the form $A (1+B\cos{2 \varphi})~$, where 
$A=0.5(N_{\rm max}+N_{\rm min})$ 
and $B=(N_{\rm max}-N_{\rm min})/(N_{\rm max}+N_{\rm min})= 
2\left< \cos{2\varphi}\right> $. 
In the model~\cite{Lokhtin:2001} the coefficient of jet azimuthal anisotropy, 
$v^{\rm jet}_{2} \equiv 
\langle \cos{2\varphi ^{\rm jet}} \rangle_{\rm event}$,  
increases almost linearly with the impact parameter $b$ and 
becomes maximum at $b \sim 1.2 R_A$. After that 
$v_2^{\rm jet}$ drops rapidly with increasing $b$: this is the domain of 
impact parameter  values  where the effect of decreasing energy loss due to 
the reduction of the  effective transverse  size of the dense zone 
and the initial energy  density of the medium is crucial and cannot be
compensated  by the stronger volume ellipticity. Anther important feature 
is  that  the jet azimuthal anisotropy decreases with increasing jet 
energy, because the energy dependence of medium-induced loss is 
rather weak (absent in the BDMS formalism and $\sim \ln E$ in the 
GLV formalism for the radiative part at high $E_T$).  

The advantage of azimuthal jet observables is that one needs to 
reconstruct only the direction of the jet, not its total energy. 
It can be  done with high accuracy, while reconstruction of the jet energy 
is more ambiguous. However, analysis of jet production as a function 
of the azimuthal angle  requires event-by-event measurement of the angular 
orientation of the reaction plane. The methods summarized in 
Ref.~\cite{Voloshin:1996,Ollitrault:1997di,Poskanzer:1998} present ways 
for reaction plane  determination. 
They are applicable for studying anisotropic particle flow in current 
heavy ion dedicated experiments at  the SPS and RHIC, and may be also used 
at the LHC~\cite{Lokhtin:2001}. Recently a method for measuring  jet azimuthal 
anisotropy coefficients without event-by-event reconstruction of the 
reaction plane was proposed~\cite{Lokhtin:2002}. 
This technique is based on the correlations between 
the azimuthal position of jet axis and the angles of hadrons not 
incorporated in the jet. The method has been generalized by taking as 
weights the particle momenta or the energy deposition in the calorimetric 
sectors. It was shown that the accuracy of the method improves with 
increasing multiplicity and particle (energy) flow azimuthal anisotropy, 
and is  practically independent of the absolute values of azimuthal 
anisotropy of the jet itself. 

\section{Conclusions} 

The azimuthal anisotropy of high-$p_T$ hadron production in
non-central heavy ion collisions is shown to  provide a 
valuable experimental tool  for studying  both gluon bremsstrahlung
in non-Abelian media and the properties of the reaction volume such 
as its size, shape, initial parton (number and energy) rapidity  densities, 
and their subsequent dynamical evolution. The {\em saturation} and  
the {\em gradual decrease} at large transverse momentum  of the reaction 
geometry generated $v_2$, predicted as a signature complementary 
to jet quenching of strong  radiative energy loss in a dense QCD 
plasma~\cite{Gyulassy:2000gk}, seem now supported by preliminary 
data extending up to $p_T \simeq 10$~GeV at RHIC.             

The initial gluon densities in $Pb+Pb$ reactions at $\sqrt{s}_{NN}=5.5$~TeV   
at the Large Hadron Collider are expected to be significantly higher
than at RHIC, implying even stronger  partonic 
energy loss. This may result in interesting novel features of jet 
quenching, such as modification of the jet distribution 
over impact parameter~\cite{Lokhtin:2000} in addition to the  
azimuthal anisotropy of the jet spectrum. The predicted large cross 
section for hard jet production on a scale of $E_T \sim 100$ GeV will 
allow for a systematic study of the differential  nuclear geometry related 
aspects of jet physics at the LHC.   

\section{Acknowledgements} The authors are grateful to J.Y.~Ollitrault for 
helpful discussion and proof reading this manuscript. It is pleasure to thank 
L.I.~Sarycheva and U.~Wiedemann for encouraging and interest in this work. 
We are much indebted to K.~Filimonov and P. Jacobs for comments on the 
STAR high-$p_T$ $v_2$ data. 
I.V. is supported  by the United States Department of Energy under 
Grant No. DE-FG02-87ER40371.


\begin{thebibliography}{99} 

\bibitem{Wang:1991ht} X.~N.~Wang and M.~Gyulassy,
Phys.\ Rev.\ D {\bf 44} (1991) 3501.

\bibitem{Eskola:1999fc}
K.~J.~Eskola, K.~Kajantie, P.~V.~Ruuskanen and K.~Tuominen,
Nucl.\ Phys.\ B {\bf 570} (2000) 379.

\bibitem{Back:2000gw}
B.~B.~Back {\it et al.}  [PHOBOS Collaboration],
Phys.\ Rev.\ Lett.\  {\bf 85} (2000) 3100.

\bibitem{Wang:xy}
X.~N.~Wang and M.~Gyulassy, Phys.\ Rev.\ Lett.\  {\bf 68} (1992) 1480.

\bibitem{Gyulassy:1993hr}
M.~Gyulassy and X.~N.~Wang, Nucl.\ Phys.\ B {\bf 420} (1994) 583.

\bibitem{Wang:1994fx}
X.~N.~Wang, M.~Gyulassy and M.~Plumer,
Phys.\ Rev.\ D {\bf 51}, 3436 (1995).

\bibitem{Baier:1998n} R.~Baier, Yu.L.~Dokshitzer, A.H.~Mueller and  D.~Schiff, 
Nucl.\ Phys.\ B {\bf 531} (1998) 403.

\bibitem{Gyulassy:2000fs}
M.~Gyulassy, P.~Levai and I.~Vitev,
Phys.\ Rev.\ Lett.\  {\bf 85} (2000) 5535.

\bibitem{Wang:2000fq}
X.~N.~Wang, Phys.\ Rev.\ C {\bf 63} (2001) 054902.

\bibitem{Gyulassy:2000gk}
M.~Gyulassy, I.~Vitev and X.~N.~Wang,
Phys.\ Rev.\ Lett.\  {\bf 86} (2001) 2537.

\bibitem{Voloshin:1996} S.A.~Voloshin and Y.~Zhang, Z.\ Phys.\ C {\bf 70} 
(1996) 665. 

\bibitem{Ollitrault:bk}
J.~Y.~Ollitrault,
Phys.\ Rev.\ D {\bf 46} (1992) 229.

\bibitem{Ollitrault:ba}
J.~Y.~Ollitrault,
Phys.\ Rev.\ D {\bf 48} (1993) 1132.

\bibitem{Kolb:2001qz}
P.~F.~Kolb, U.~W.~Heinz, P.~Huovinen, K.~J.~Eskola and K.~Tuominen,
Nucl.\ Phys.\ A {\bf 696} (2001) 197.

\bibitem{Teaney:2001av}
D. Teaney, J. Lauret, and  E.V. Shuryak, nucl-th/0110037. 

\bibitem{Molnar:2001ux}
D.~Molnar and M.~Gyulassy, Nucl.\ Phys.\ A {\bf 697} (2002) 495
[Erratum-ibid.\ A {\bf 703} (2002) 893].

\bibitem{Lin:2001zk}
Z.~W.~Lin and C.~M.~Ko, Phys.\ Rev.\ C {\bf 65} (2002) 034904.

\bibitem{Shuryak:2001me}
E.~V.~Shuryak, Phys.\ Rev.\ C {\bf 66} (2002) 027902.

\bibitem{Teaney:2002kn}
D.~Teaney and R.~Venugopalan, Phys.\ Lett.\ B {\bf 539} (2002) 53.

\bibitem{Kovchegov:2002nf}
Y.~V.~Kovchegov and K.~L.~Tuchin, Nucl.\ Phys.\ A {\bf 708} (2002) 413. 

\bibitem{Wang:qh}
S.~Wang {\it et al.}, Phys.\ Rev.\ C {\bf 44} (1991) 1091.

\bibitem{Borghini:2001vi}
N.~Borghini, P.~M.~Dinh and J.~Y.~Ollitrault,
Phys.\ Rev.\ C {\bf 64} (2001) 054901.

\bibitem{Borghini:2001zr}
N.~Borghini, P.~M.~Dinh and J.~Y.~Ollitrault, nucl-ex/0110016.

\bibitem{Gyulassy:1999ig}
M.~Gyulassy, P.~Levai and I.~Vitev, Nucl.\ Phys.\ A {\bf 661} (1999) 637

\bibitem{Gyulassy:1999zd}
M.~Gyulassy, P.~Levai and I.~Vitev, Nucl.\ Phys.\ B {\bf 571} (2000) 197

\bibitem{Gyulassy:2000er}
M.~Gyulassy, P.~Levai and I.~Vitev, 
Nucl.\ Phys.\ B {\bf 594} (2001) 371.

\bibitem{Levai:2001hd}
P.~Levai, G.~Papp, G.~Fai and M.~Gyulassy,
J.\ Phys.\ G {\bf 28} (2002) 2059.

\bibitem{Filimonov:2002xk}
K.~Filimonov  [STAR Collaboration], arXiv:nucl-ex/0210027.

\bibitem{Jacobs:2002pz}
P.~Jacobs, arXiv:hep-ex/0211031.

\bibitem{Vitev:2002pf}
I.~Vitev and M.~Gyulassy, arXiv:hep-ph/0209161, Phys. Rev. Lett. in press.

\bibitem{Vitev:2001zn}
I.~Vitev and M.~Gyulassy, Phys.\ Rev.\ C {\bf 65} (2002) 041902.

\bibitem{Lokhtin:2000} I.P.~Lokhtin and A.M.~Snigirev, Eur.\ Phys.\ J.\ C {\bf 16} 
(2000) 527. 

\bibitem{pythia} T.~Sj\"ostrand, Comput.\ Phys.\ Commun.\ {\bf 82} (1994) 74. 

\bibitem{Vogt:1999} R.~Vogt, Heavy Ion Phys.\ {\bf 9} (1999) 339. 

\bibitem{Lokhtin:2001} I.P.~Lokhtin,  S.V.~Petrushanko, L.I.~Sarycheva and 
A.M.~Snigirev, [arXiv:hep-ph/0112180], Proceeding of International 
Conference on Physics and Astrophysics of Quark-Gluon Plasma (Jaipur, India, 
26-30 Nov 2001); Phys.\ At.\ Nucl.\ {\bf 65} (2002) 943.    

\bibitem{Bjorken:1983} J.D.~Bjorken, Phys.\ Rev.\ D {\bf 27} (1983) 140.  

\bibitem{Eskola:1994} K.J.~Eskola, K.~Kajantie and P.V.~Ruuskanen, Phys.\ Let.\ B 
{\bf 332} (1994) 191. 

\bibitem{Eskola:1997} K.J.~Eskola, Prog.\ Theor.\ Phys.\ Suppl.\ {\bf 129} 
(1997) 1. 

\bibitem{Eskola:1998} K.J.~Eskola, Comments Nucl.\ Part.\ Phys.\ {\bf 22} 
(1998) 185. 

\bibitem{Lokhtin:1998} I.P.~Lokhtin and A.M.~Snigirev, Phys.\ Lett.\ B {\bf 163} (1998) 
440.

\bibitem{Baier:1998} R.~Baier, Yu.L.~Dokshitzer, A.H.~Mueller and D.~Schiff, 
Phys.\ Rev.\ C {\bf 58} (1998) 1706.

\bibitem{Damgov:2001} I.~Damgov {\it et al.}, Part.\ Nucl.\ Lett.\ 
{\bf 107} (2001) 93; CERN CMS Note 2001/055. 

\bibitem{Ollitrault:1997di}
J.~Y.~Ollitrault, arXiv:nucl-ex/9711003.

\bibitem{Poskanzer:1998} A.M.~Poskanzer and S.A.~Voloshin, Phys.\ Rev.\ C 
{\bf 58} (1998) 1671.

\bibitem{Lokhtin:2002} I.P.~Lokhtin, L.I.~Sarycheva and A.M.~Snigirev, 
Phys.\ Lett.\ B {\bf 537} (2002) 261.

\end{thebibliography}
\end{document}